\documentstyle [12pt] {article}

\catcode`@=12
\newcommand{\be}{\begin{equation}}
\newcommand{\ee}{\end{equation}}
\newcommand{\ber}{\begin{eqnarray}}
\newcommand{\eer}{\end{eqnarray}}
\newcommand{\bers}{\begin{eqnarray*}}
\newcommand{\eers}{\end{eqnarray*}}
\begin{document}
\vspace{0.5in}
\oddsidemargin -.375in  
\newcount\sectionnumber 
\sectionnumber=0 
\def\be{\begin{equation}} 
\def\ee{\end{equation}}
\thispagestyle{empty}  
\begin{flushright} AMES-HET-96 \\UH-511-852-96\\June 1996\
\end{flushright}
\vspace {.5in} 
\begin{center} 
{\Large\bf Pion-Nucleon Phase Shifts in\\ Heavy Baryon Chiral
Perturbation Theory \\}
\vspace{.5in} 
{\bf Alakabha Datta${}^{a)}$} and 
{\bf Sandip Pakvasa${}^{b)}$ \\}
\vspace{.1in} 
${}^{b)}$ {\it Department of Physics and Astronomy, Iowa State 
University, Ames, Iowa.}\\  
${}^{b)}$ {\it 
Department of Physics and Astronomy, University of Hawaii at Manoa, 
Honolulu, HI 96822, USA.}\\  

\end{center}  

\begin{abstract}  
We calculate the phase shifts in the pion-nucleon scattering using the
heavy baryon formalism. We consider phase shifts  for the 
pion  energy range of
$140$ to $200 $ MeV. We employ two different methods for calculating the
phase shifts - the first using the full third order calculation of the
pion-nucleon scattering amplitude and the second by including the
resonances
$\Delta$ and $N^*$ as explicit degrees of freedom in the Lagrangian. We
compare the results of the two methods with phase shifts
extracted from fits to the pion-nucleon scattering data. We find
good to fair agreement between the calculations and the phase shifts from
 scattering data.
\end{abstract}  
\newpage
\baselineskip 24pt
 The amplitudes for non-leptonic decays of hyperons are modulated by the
final state strong scattering a la the Fermi-Aidzu-Watson
theorem\cite{Fermi}. These final state phase shifts are crucial in
calculating the various CP violating asymmetries in hyperon decays
\cite{pak}. Some of the asymmetries depend on $\sin \delta$ where
$\delta$ is some combination of the final state scattering phase shifts
and a knowledge of $\delta$ is necessary to make predictions for CP
violations in hyperon decays. Recently there have been new calculations 
\cite{ml}
of $\Lambda-\pi$ scattering phase shifts in the framework of heavy
baryon chiral perturbation theory (HBCHPT), with much smaller results
than some earlier dispersive estimates \cite{Rnath}.
 Calculations of $\Lambda-\pi$ phase shifts are 
relevant to the measurement of CP violation in the hyperon 
decay $\Xi \rightarrow \Lambda \pi$ \cite{pak}. An experiment to measure the
combined asymmetry $ \Delta\alpha = \Delta\alpha_{\Lambda} +
\Delta\alpha_{\Xi} $ will be carried out in the near future at Fermilab
\cite{fl}. Here for example; $ \Delta\alpha_{\Xi} = \alpha_{\Xi} +
\bar{\alpha_{\Xi}}$ and $ \Delta\alpha_{\Xi} $ is proportional to $
\tan(\delta_s-\delta_p) $. As a test of how reliable 
 these phase shift calculations might be, we apply the same techniques
to calculate the pion-nucleon phase shift in the energy range
$E_{\pi}=140
-200$ MeV,
where experimental data exist and can be compared to the predictions. It
is important to point out that the calculations in Ref\cite{ml} were done
in a $SU(2)_L\times SU(2)_R$ HBCHPT and so we believe that a calculation
in the $\pi-N$ sector should be a good test of the reliability of 
the calculations in
Ref\cite{ml}.

The pion-nucleon Lagrangian in the heavy baryon has been written down to
$0(p^3)$ ( See Ref \cite{Meissner} and the references there in.) 
and has been used, to calculate the S-wave
scattering lengths for pion nucleon scattering
 at threshold \cite{Meissner2}. Recently the pion nucleon scattering has been
calculated to $0(p^3)$ in Ref\cite{Mo}.
 We will use the
Lagrangian to calculate the phase shifts away from threshold. Our
purpose in this paper is to check that calculations of the phase shifts based
on the HBCHPT are in reasonable agreement with the experimental data. 
We will not be aiming for precise matching of the theoretical and the
experimental numbers but rather we will be satisfied if our calculations
agree with the data to a factor of about two. In the next
section we describe the pion-nucleon Lagrangian and describe our
calculations while in the following section we discuss our results and 
summarize our calculations.

\section{Pion-Nucleon Lagrangian}
The basic framework that we will employ to calculate the phase shifts is
the heavy baryon chiral perturbation theory \cite{me}. As is well known, the
relativistic formalism of the chiral Lagrangian with baryons suffers
from some severe problems which arise because the baryon mass does not
vanish in the chiral limit and is not small relative to the scale of
chiral symmetry breaking scale $\Lambda_{\chi}$. The 
 baryon four momentum is, therefore,
not small relative to $\Lambda_{\chi}$ which results in the loss of the
one to one correspondence between the loop and the small momentum expansion
in CHPT.
 In the heavy baryon
formalism, the baryons are treated as static sources and only the baryon
momentum relative to the rest mass is important. The Lagrangian in this
limit is constructed by taking the the extreme non-relativistic limit
of the relativistic Lagrangian and expanding in inverse powers of the
heavy baryon masses. In this formalism
 the  one to one correspondence between the loops and the
small momentum expansion is restored.

The lowest order $SU(2)_L\times SU(2)_R$ invariant pion nucleon 
Lagrangian can be written as \cite {Meissner}
\ber
{\cal{L}}_{\rm relativistic} & = &
	\bar{\psi} \, [i\gamma \cdot D -m+\frac{g_A}{2}
 \gamma^{\mu} \gamma_5 u_\mu] 
	\,\psi \nonumber\\
& \stackrel{\rm static \ limit}{\longrightarrow} & 
	\bar{\psi} \, [i\gamma \cdot D + g_A S\cdot u] \, \psi
\eer
where
\bers
S_{\mu} & = & \frac{i\gamma_5\sigma_{\mu\nu} v^{\nu}}{2}\\ 
D_{\mu} & = & \partial_{\mu} + V_{\mu}  \\
V_{\mu} & = & \frac{1}{2} {[\, \xi^{\dag} \partial^{\mu} \xi 
	+ \xi \partial^{\mu} \xi^{\dag}]} \\
u_{\mu} & = & {i} {[\, \xi^{\dag} \partial^{\mu} \xi 
	- \xi \partial^{\mu} \xi^{\dag}]} \\
\xi & = & e^{\frac{i\pi_a T_a}{F_{\pi}}}, \hspace{4cm} 
	( \pi_a{\quad} {\rm{represents \ the \ pion \ with \ isospin \
index \ a}} )\
\eers
$v^{\nu}$ is the nucleon four velocity and in the rest frame of the
nucleon $S_{\mu}$ is the spin operator.
$F_{\pi}=93MeV$ is the pion decay constant, $T_a=\frac{1}{2}\tau^a$ 
are the isospin 
generators and $g_A \sim 1.26$ is the axial vector coupling measured in
the neutron beta decay. 
 This Lagrangian generates the following vertices
\bers
{\rm{vector \ coupling}} & : & \frac{1}{4 F_{\pi}^2} \, \varepsilon^{abc} \, 
	(v \cdot k_1 + v \cdot k_2) \, \tau^c \\
{\rm{axial \ vector \ coupling}}& : & \frac{g_A}
{F_{\pi}} \, S \cdot k_2 \, \tau^a
\eers
 where $k_1$ is the incoming pion momentum and $k_2$ is 
the outgoing pion momentum. To the Lagrangian above we have to add the
pion Lagrangian ${\cal{L_{\pi,\pi}}}$ whose form is well known (for
a recent review see Ref\cite{Meissner}).
 We can organize the Lagrangian in terms of the small momentum in
the calculation
\begin{eqnarray}
{\cal{L}}(\pi,N) & = & {\cal{L}}^1(\pi,N) + {\cal{L}}^2(\pi,N) + 
{\cal{L}}^3(\pi,N)
\eer
where ${\cal{L}}^i(\pi, N) $ generates
terms of $ \sim p^i$, $p$ being the the small momentum. As already
mentioned the scattering amplitude has been calculated to third order in
the small momentum $p$ in Ref\cite{Mo}. The low energy constants that
appear in the chiral Lagrangian were fixed by fitting to the available
pion-nucleon data on
the threshold parameters like scattering lenghts, volume,
effective ranges, etc, the pion-nucleon $\sigma$ term and the
Goldberger - Treiman discrepancy. It was noted in Ref\cite{Mo} and also
verified by us that the full third order calculation results in a better
description of the data. However the expansion in the small momentum appears
to be slowly converging with contributions from the various orders being
sometime of the same magnitude indicating the importance of higher order
calculations. The calculations are expected to give good
agreement with data near threshold but we will use these results to
calculate the phase shifts away from threshold in the energy range
$ 1080 \stackrel{<}{=} \sqrt{s}\stackrel{<}{=}1130 $ MeV. As mentioned
in the introduction we are only intersted in a reasonable agreement, by
about a factor of two, of
our calculations with the experimental data.

The low energy constants
 in
${\cal{L}}^2(\pi,N) + {\cal{L}}^3(\pi,N)$  have also been estimated from the
resonance exchange approximation. The basic idea is to start from a fully
relativistic Lagrangian involving the pion and the nucleon resonances.
The resonances are then integrated out from the theory to produce the
higher dimensional terms in the pion-nucleon Lagrangian. In this scenario
the $\Delta$ and the $N^*$(1440) and the other resonances
 are not included as dynamical degrees
 of freedom in the
effective theory for pion-nucleon scattering. The resonance
approximation is clearly expected to work well close to threshold so
that the resonances are far enough to be integrated out of the theory.
The fact that the $\Delta(1232)$ is sufficiently close to the energy region
we are interested in can be a motivation to include the $\Delta$ as an
explicit degree of freedom in the chiral Lagrangian. 
We therefore include the $\Delta$ in our calculations and write the total
Lagrangian as
\ber
{\cal{L}} & = & {\cal{L}}(\pi, N, \Delta) \nonumber\\
    & = & {\cal{L}}(\pi, N) + {\cal{L}}(\pi, N, \Delta) \
\eer
 The disadvantage of including 
 the $\Delta$ as an explicit degree of freedom is that the consistent 
power counting
in HBCHPT is destroyed. This problem can be solved by treating
$m_{\Delta}-m_N=\Delta$ as a small parameter and then consider a chiral
Lagrangian expansion in the small momentum $p$ which now includes the
$\Delta$\cite{Hol}. Moreover the full third order calculation should
also include the additional loop contributions due to the $\Delta$.
 In this paper, however, we will not follow the
approach of Ref\cite{Hol} but for the purpose of the paper it will be
sufficient to include only the tree level contribution from
${\cal{L}}(\pi, N) + {\cal{L}}(\pi, N, \Delta) $. A calculation based on
this approximation may give a better description of the data for the
$P_{33}$ phase shift, where the $\Delta$ shows up as a resonance, compared
to a calculation without the $\Delta$ included in the Lagrangian.
However we would like to stress that a full consistent third order calculation
including the $\Delta$ should be performed before drawing any conclusion
about the inclusion of the $\Delta$ in the Lagrangian. We will also
include the $N^*(1440)$ in the Lagrangian whose effect may be important
for the $P_{11}$ phase shift. We will
also expand the tree level contribution of the
$\Delta$ and the $N^*$ resonances in powers of the small momentum and
try to understand the dominant contributions \cite{Meissnernew1}.

Following
 Jenkins and Manohar \cite{me2} one can incorporate the $\Delta$ by writing
the Lagrangian
\ber
{\cal{L}}({\pi N \Delta}) & = & \frac{3g_A}{2\sqrt{2} F_{\pi}} 
	[\bar{T}^{\mu a} u{_\mu}^a N 
	+ \bar{N} u_{\mu}^a T^{\mu a}] \\
T^{\mu 1} & = & \frac{1}{\sqrt{2}}
	\left[ \begin{array}{c}
	\Delta^{++} - \frac{\Delta^0}{\sqrt{3}} \\
	\frac{\Delta^+}{\sqrt{3}} - \Delta^-
	\end{array} \right] \\
T^{\mu 2} & = & \frac{i}{\sqrt{2}}
	\left[ \begin{array}{c}
	\Delta^{++} + \frac{\Delta^0}{\sqrt{3}} \\
	\frac{\Delta^+}{\sqrt{3}} + \Delta^-
	\end{array} \right] \\
T^{\mu 3} & = & -\sqrt{\frac{2}{3}}
	\left[ \begin{array}{c} \Delta^+ \\ \Delta^0 \end{array} \right]
\eer
where
\bers
u_{\mu}^a & = & iTr(T^a \xi^+\partial_{\mu}U\xi^+) \\
U & = &\xi^2 \\
\eers
 We can include the $N^*$(1440) in our calculation also and write the
interaction Lagrangian involving the $N^*$, pion and the nucleon as
\ber
{\cal{L}}({\pi N N^*}) &= &g_{N^* N \pi}\bar{\psi_{N^*}} \,  S\cdot u \,
\psi_{N}\
\eer
The coupling $g_{N^* N \pi}$ can be fixed from the branching fraction of
$N^*\rightarrow N \pi$ \cite{PDG}.

In the static limit the momentum of off shell $N, \Delta$
and $N^*$(1440) are written as,
\bers
P_N & = & m_N v + k \\
P_\Delta & = & m_\Delta v + k\\
P_{N^*} & = & m_{N^*} v + k\
\eers
where
\bers
k &\sim & p \sim k_{\pi}\ll  m_{N,\Delta, N^*} \
\eers

 The propagators for the various fields in the heavy baryon limit are
\bers
S_F(N) & = & \frac{i}{v \cdot k}=S_F(N^*) \\
S_F(\Delta)& =& -\frac{i}{v \cdot k -\Delta} 
	\left[ g^{\mu\nu} - v^\mu v^\nu + \frac{4}{3} S^\mu S^\nu \right]\
\eers
while the pion propagator is
\bers
\Delta(\pi) & = & \frac{i}{q^2 - m_\pi^2} \
\eers
\subsection{Calculation procedure}
The pion-nucleon scattering amplitude with the outgoing(incoming) pion
carrying the isospin index b(a) can be written as
\ber
T^{ba}& =& T^{+} \delta^{ab} + i\varepsilon^{bac} T^{-} \tau^c
\eer
Our job is to calculate  $T^{+}, T^{-}$. Having obtained these we can
isolate the $T(\rm isospin)=\frac{1}{2}$ and
$T(\rm isospin)=\frac{3}{2}$ amplitudes in the following manner. We first
calculate
\bers
T ({p\pi^-} \rightarrow p\pi^-) &= & T^+ + T^{-} \\
T ({p\pi^-} \rightarrow n\pi^0) & = & -\sqrt{2} \, T^{-}
\eers
We can then use these amplitudes to extract
\ber
a_{T=\frac{1}{2}} & = & T(p\pi^- \rightarrow p\pi^-)  
	- \frac{1}{\sqrt{2}} \, T(p\pi^- \rightarrow n\pi^0) =T^{+} +2T^{-}  \\
a_{T=\frac{3}{2}} & = & T(p\pi^- \rightarrow p\pi^-) 
	+ \sqrt{2} \, T(p\pi^- \rightarrow n\pi^0) = T^{+} -T^{-}  \
\eer

Next we expand the isospin amplitudes in partial waves ($S, P$ waves) as
\ber
a_{T=\frac{1}{2}} & = & S_{11} + (2P_{13} + P_{11})\cos\theta
	+ i\sin\theta \, {\bf \sigma} \cdot \hat{\bf n} \, (P_{13} - P_{11}) \\
a_{T=\frac{3}{2}} & = & S_{31} + (2P_{33} + P_{31})\cos\theta
	+ i\sin\theta \, {\bf \sigma} \cdot \hat{\bf n} \, (P_{33} - P_{31})
\eer
where the subscripts refer to twice the 
 $I({\rm{isospin}})$ and $J({\rm{angular \ momentum}})$ values,
$\hat{\bf n}=\frac{{\bf k_1}\times{\bf k_2}}
{|{\bf k_1}\times{\bf k_2}|}$ and $\theta$ is the scattering angle. 

 For a given partial wave amplitude $A_{2I,2J}$ the phase shift
$\delta_{2I,2J}$, for small phase shift, is
\ber
{A_{2I,2J1}} \sim {e^{i\delta_{2I,2J}}\sin\delta_{2I,2J}} \approx 
 \delta_{2I,2J}
\eer 
where $s$ is the c.m energy. The scattering amplitude is in general
complex and it develops an imaginary part at order $p^3$. The above
approximation assumes that the imaginary part of the scattering
amplitude is small for small phase shifts.
 The contribution from ${\cal{L}}^{1}$ to the amplitudes are
\ber
T^{+} &=&\frac{g_A^2}{F_{\pi}^2 \omega}
i\varepsilon^{\mu\nu\rho\sigma} v_\rho S_\sigma k_{1\mu} k_{2\nu}
\nonumber\\
T^{-} &= &
\frac{\omega}{2F_{\pi}^2}\left[1-\frac{g_A^2(\omega^2 -k_1
\cdot k_2)}{\omega^2}
\right]\
\eer
where $\omega = v \cdot k$. The 
 contributions from  the $\Delta$ exchange are
\be
T(p\pi^- \rightarrow p\pi^- {\rm with \ \Delta \ exchange})
	= \left[\frac{3g_A}{2\sqrt{2}F} \frac{2}{3}\right]^2 i\bar{u}_p 
	\left[(\omega^2 -k_1 \cdot k_2)
{\frac{\omega -2\Delta}{\omega^2 - \Delta^2}} {} + 
	i\varepsilon^{\mu\nu\rho\sigma} v_\rho S_\sigma k_{1\mu} k_{2\nu}
	{\frac{\Delta - 2\omega}{\omega^2 - \Delta^2}}\right] u_p
\ee
\ber
T(p\pi^- \rightarrow n\pi^0) = \left[\frac{3g_A}{2\sqrt{2}F} 
	\frac{2}{3} \right]^2 \sqrt{2} i(-\bar{u}_n)
	\left[(\omega^2 - k_1\cdot k_2) \frac{\omega}{\omega^2 - \Delta^2}
	+ i\varepsilon^{\mu\nu\rho\sigma} v_\rho S_\sigma k_{1\mu} k_{2\nu}
	\frac{\Delta}{\omega^2 -\Delta^2} \right] u_p \
\eer
where $$ \Delta = m_\Delta - m_N.$$
From these expressions one can check that there is a pole only for the
$P_{33}$ channel where the $\Delta$ shows up as a resonance.

 Following Ref\cite{Meissnernew1} we can, for $\omega<\Delta $, 
expand the $\Delta$ contribution in powers of $\omega/\Delta$. We can
think of the various terms in the expansion as coming from higher order
terms, beginning at order $p^2$, of the chiral Lagrangian. We will
truncate the expansion at order $p^4$ and compare the results with the 
calculation done with including the full $\Delta $ contribution. 

The contributions from $N^*$ are
\ber
T^{+} &=&\frac{g_{N^* N \pi}^2}{F_{\pi}^2 \omega}
i\varepsilon^{\mu\nu\rho\sigma} v_\rho S_\sigma k_{1\mu} k_{2\nu}
\frac{\omega}{\omega^2 -M^2} - \frac{g_{N^* N \pi}}{2F_{\pi}^2}
(\omega^2 -k_1 \cdot k_2) \frac{M}{\omega^2 - M^2}
\nonumber\\
T^{-} &= &
\frac{g_{N^* N \pi}^2}{F_{\pi}^2 \omega}
i\varepsilon^{\mu\nu\rho\sigma} v_\rho S_\sigma k_{1\mu} k_{2\nu}
\frac{M}{\omega^2 -M^2} - \frac{g_{N^* N \pi}}{2F_{\pi}^2}
(\omega^2 -k_1 \cdot k_2) \frac{\omega}{\omega^2 - M^2}\
\eer 
where $$ M = m_{N^*}- m_N$$
In this case the pole shows up in the $P_{11}$ channel
at the $N^*$ mass.
Like in the case for the $\Delta$, we will also do an expansion to order
$p^4$ by expanding the denominator in the above equations in powers of
$\omega/M$.

We therefore consider two
cases
in our calculations. In the first case we include the $\Delta$
 and the $N^*$(1440) in the
effective Lagrangian and consider only the tree level contributions. For the
second case we will use the full third order calculation without the
$\Delta$ and the $N^*$ given in Ref\cite{Mo}. All the details of the
calculation can be found in Ref\cite{Mo} and we do not repeat them here.
We will call the results of the calculations of the two cases as Result
1 and Result 2.

\section{Results and Discussions}
In this section we present and discuss our results. We show our
results for the $S_{11}$, $S_{31}$, $P_{11}$, $P_{13}$, $P_{31}$ and
$P_{33}$
 partial waves. We
compare our result to phase shifts extracted from fits to pion nucleon
scattering data which were obtained from the SAID program \cite {SAID}.
 In each figure we show generally three curves. Two curves
shows the calculation for the two cases described above and the
the third curve
shows the phase shifts obtained from fits to experimental data from the
SAID program. The errors for the phase shifts are available from the SAID
program
for certain single energy values. They vary typically from $0.1$ to $0.3$
degrees. In the energy range we are interested in there are no error 
calculations available for the phase shifts for the $P_{13}$ and the
$P_{31}$
partial waves. We, therefore, do not show the error bars in our graphs.
 We also limit ourselves to small phase shifts, typically,
$< 10$ degrees or in the energy range $1080 - 1130$ MeV corresponding
to a pion energy range $E_{\pi}\approx 140-200 $ MeV which includes the
$\Lambda
\rightarrow N \pi $ region.

 In Fig. 1 we show the $S_{11}$ phase shifts. We find the agreement with data
is good to about $1100$ MeV for calculation 1 and 2
  beyond which the calculated phase shifts
are
larger than the experimental ones. 
The results
of calculation 2 based on the calculations of Ref\cite{Mo} are in 
better agreement with the data compared to
calculation 1. 
In Fig. 2 we show the $S_{31}$ phase shifts. The agreement between the results
of calculation 1 and experiment is reasonably good with the result
of calculation 1 being in better agreement with data then calculation 2 though
near threshold the result of calculation 2 is in better agreement with data.
In Fig. 3 we show the $P_{11}$ phase shifts. The agreement between data
and the calculation is better for Result 2 which
describes the data quite well  for most of the energy
range. The tree level contribution from the $N^*$ included in Result 1
improves agreement with data but as already mentioned
 the result of calculation 2 describes the data better
than calculation 1 indicating perhaps the importance of the other
resonances whose effects are included in the low energy constants of the
chiral Lagrangian if one believes in the resonance approximation. We
also note that the $N^*(1440)$ may be far enough in mass from our region
of interest to give the dominant contribution to the phase shift. 
In Fig. 4 we show the $P_{13}$ phase shifts. There is
moderate agreement with data. The agreement of calculation 2
 with data is much better than that of calculation 1 again.
In Fig. 5 we show the $P_{31}$ phase shifts. Like $P_{13}$ the agreement is
moderately good
with data with results of calculation 1 and calculation 2  being
less and more than the data. The result of calculation 2 is in good
agreement with the data to the centre of mass energy of 1100 MeV.
In Fig. 6 we show the $P_{33}$ phase shifts. The agreement here with
experiment is quite good. In this case calculation 1
is in better agreement with experiment than the results
of calculation 2. We also
perform calculation 1 but expand the $\Delta$ and $N^*$
contributions to $O(p^4)$. The  result of this calculation
is close to those of calculation 1
 to about $E_{cm}=1100 $ Mev, where the the effects of the higher
order terms ( higher than $p^4$) in the expansion are not significant.
Similar
 agreements
between the results of this  calculation and calculation 1
 are also found for the
other P and the S wave phase shifts for
$E_{cm}\stackrel{<}{=}1100 $ MeV. 

In the light of our results it appears that the results
 of calculation 2 are in better agreement with data than
the results of calculation 1 except for the $S_{31}$ and the $P_{33}$
phase shifts. For the $P_{33}$ phase shift this is probably due to the fact the
$\Delta$ contribution dominates the phase shift. However as metioned
earlier the expansion in small momentum is slowly converging and higher
order terms are important. The inclusion of higher order terms in
calculation 2  may improve the agreement with data for the $P_{33}$
phase shift. 

 In summary we have calculated the S and P phase shifts in pion nucleon
 scattering in heavy baryon chiral perturbation theory. We find that
HBCHPT can give reasonable description of the phase shifts away from
threshold in the enegy range $1080-1130$ MeV. The inclusion of the $\Delta$  
may be necessary for the $P_{33}$ phase shift but no final conclusion
can be reached until higher order calculations are done.
 Based on these results, the results obtained earlier for
$\Lambda-\pi$ phase shifts, are probably correct 
within a factor of two and thus confirm the smallness of the
$\Lambda-\pi$ phase shifts and in turn the smallness of
$\Delta\alpha_{\Xi}$. 

{\bf Acknowledgment:}
We would like to thank G. Valencia, B.L. Young and K. Lassila
for useful discussions. 
This work was supported in part  by US 
D.O.E grant \# DE-FG 03-94ER40833 (S. Pakvasa) and \# DE-FG02-92ER40730 (A.
Datta).

\subsection{Figure Captions}
\begin{itemize}

\item {\bf Fig .1}: $S_{11}$ phase shifts. 
The solid, dotted lines
 correspond to calculation 1, 2 while the symbol 'circle'
represent the phase shifts extracted from fits to the pion nucleon
scattering data.

\item {\bf Fig .2}: $S_{31}$ phase shifts . The solid, dotted lines
 correspond to calculation 1, 2 while the symbol 'circle'
represent the phase shifts extracted from fits to the pion nucleon
scattering data.

\item {\bf Fig .3}: $P_{11}$ phase shifts. The solid, dotted lines
 correspond to calculation 1, 2 while the symbol 'circle'
represent the phase shifts extracted from fits to the pion nucleon
scattering data.

\item {\bf Fig .4}: $P_{13}$ phase shifts. The solid, dotted lines
 correspond to calculation 1, 2 while the symbol 'circle'
represent the phase shifts extracted from fits to the pion nucleon
scattering data. 

\item {\bf Fig .5}: $P_{31}$ phase shifts. The solid, dotted lines
 correspond to calculation 1, 2 while the symbol 'circle'
represent the phase shifts extracted from fits to the pion nucleon
scattering data.

\item {\bf Fig .6}: $P_{33}$ phase shifts. The solid, dotted lines
 correspond to calculation 1, 2 while the symbol 'circle'
represent the phase shifts extracted from fits to the pion nucleon
scattering data. In this figure we also show the result of calculation 1
( Result 3) with the $\Delta$ and the $N^*$ contributions expanded to
$O(p^4)$
and is represented by the long-dashed line.

 \end{itemize}


\begin{thebibliography}{References}
\bibitem{Fermi} E. Fermi, Suppl. Nuovo Cimento, {\bf 2}, 58 (1955);
K.M. Watson, Phys. Rev. {\bf 95}, 228 (1954); K. Aidzu, Proc. of
International Conf. on Theoretical Physics, (1953), Kyoto-Tokyo Science
Council, Japan(1954), p.200; A detailed discussion as well as 
a complete list of
references can be found in
 Final State Interactions, J. Gillespie, 
San Francisco, Holden-Day Publishers (1964).

\bibitem{pak} J. Donoghue, Xiao-Gang He and S. Pakvasa, Phys. Rev. {\bf
D34}, 833 (1986); 
 Xiao-Gang He and S. Pakvasa, UH-511-802-94, August (1994).

\bibitem{ml} M. Lu, M. Wise and M. Savage, Phys. Lett. {\bf B 337}, 133
(1994); A. Datta and S. Pakvasa, Phys. Lett. {\bf B 344}, 430 (1995).


\bibitem{fl} J. Antos {\it et al.}, Fermilab Proposal P-871 (revised
version), March 26, 1994.


\bibitem{Rnath} B.R. Martin, Phys. Rev. {\bf 138}, 1136 (1965);
 R. Nath and A. Kumar, Nuovo Cimento {\bf 36}, 669 (1965).

\bibitem{Meissner} V. Bernard, N. Kaiser and U.-G. Mei\ss ner,
Int. J. Mod. Phys. {\bf E 4}, 193 (1995); V. Bernard, N. Kaiser and 
U.-G. Mei\ss ner,
Nucl.Phys. {\bf A 615} 483 (1997).

\bibitem{Meissner2} V. Bernard, J. Gasser, N. Kaiser and U.-G. Mei\ss ner,
Phys. Lett. {\bf B 268}, 291 (1991); V. Bernard, N. Kaiser 
and U.-G. Mei\ss ner,
Phys. Rev. {\bf C 52}, 2185 (1995).

\bibitem{Mo} Martin Moj\v{z}i\v{s} hep-ph/9704415.

\bibitem{me} E. Jenkins and A.V. Manohar, Phys.Lett. {\bf B 255}, 558 
(1991); E. Jenkins and A.V. Manohar, in Proceedings of the Workshop 
on Effective Field
Theories of the Standard Model, Debogoko, Hungary,Aug. 22-26 (1991),ed. U.-G.
Mei\ss ner, World scientific.

\bibitem{Hol} Thomas R. Hemmert, Barry R. Holstein and Joachim Kambor,
Phys.Lett. {\bf B 395}, 89 (1997).

\bibitem{Meissnernew1} V. Bernard, N. Kaiser and U.-G. Mei\ss ner, Z.
Phys {\bf C 60}, 111 (1993).

\bibitem{me2} E. Jenkins and A.V. Manohar, Phys.Lett. {\bf B 259}, 353
(1991).

\bibitem{PDG} Particle Data Group, Phys.Rev. {\bf D 50}, 1223 (1994).

\bibitem{SAID} Scattering Analyses Interactive Dial-In program code. The
experimental data file used was PN961f PI-N data VPI and ISU, Arndt 02/26/96.
\end{thebibliography}
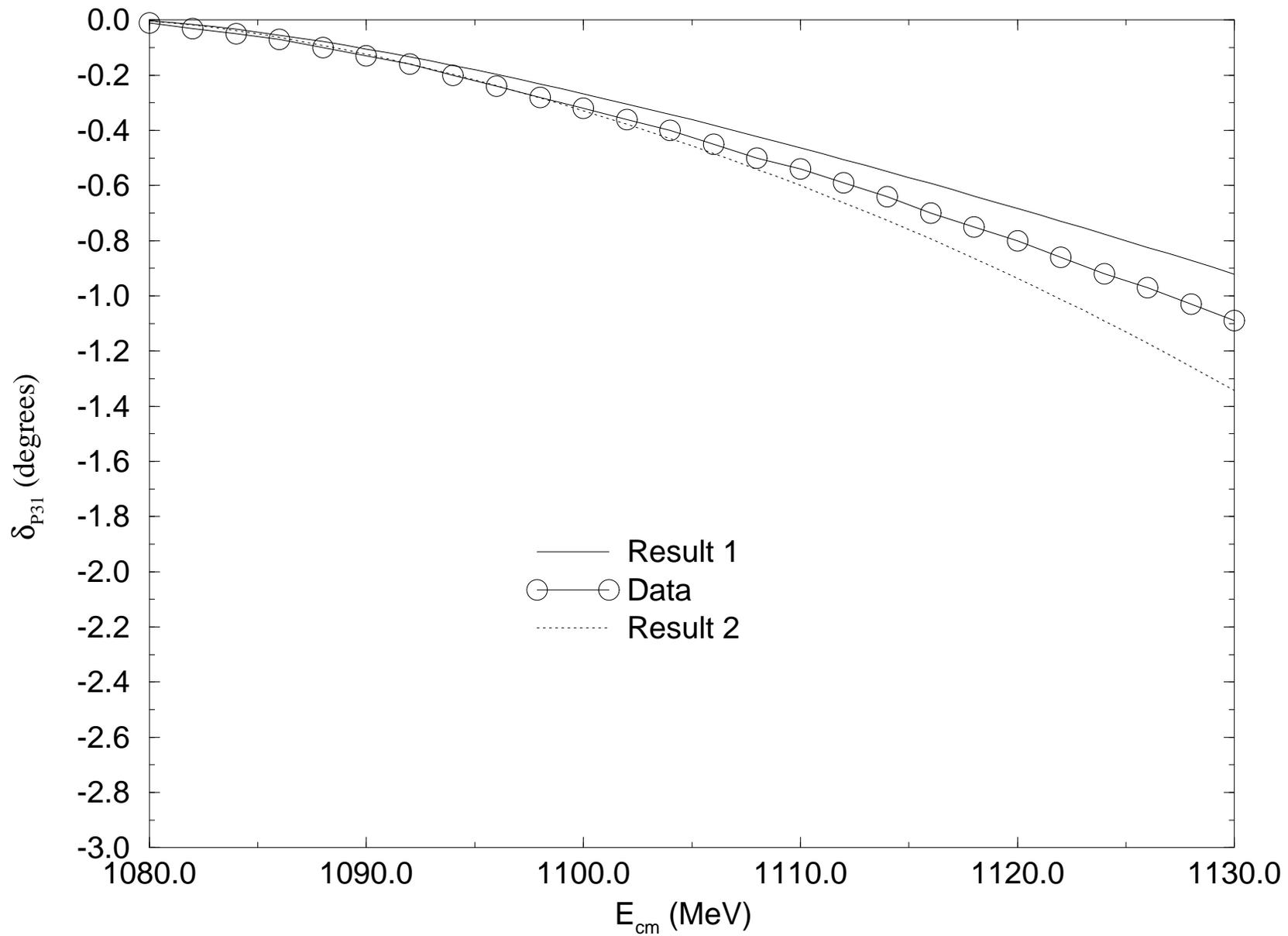
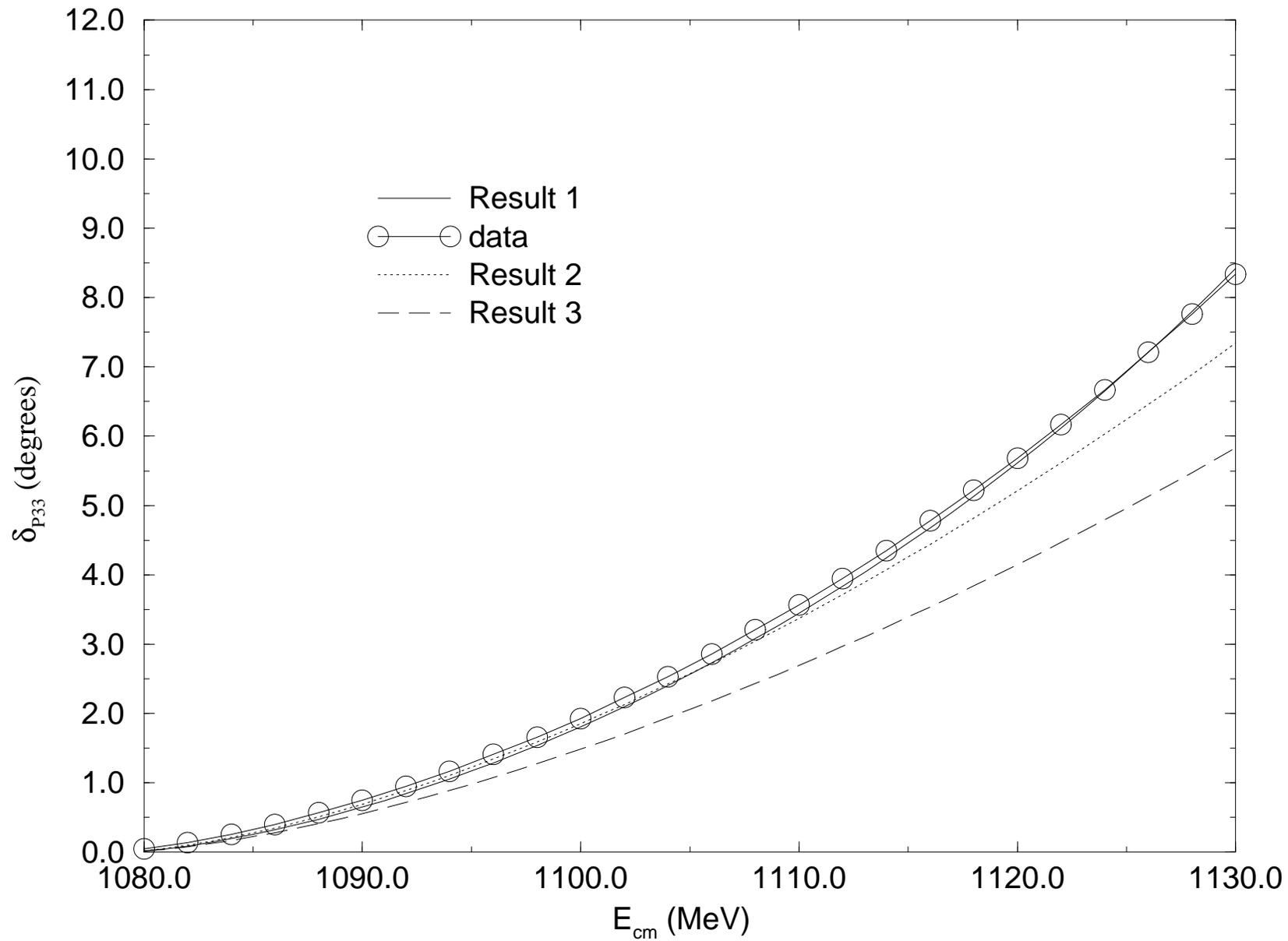
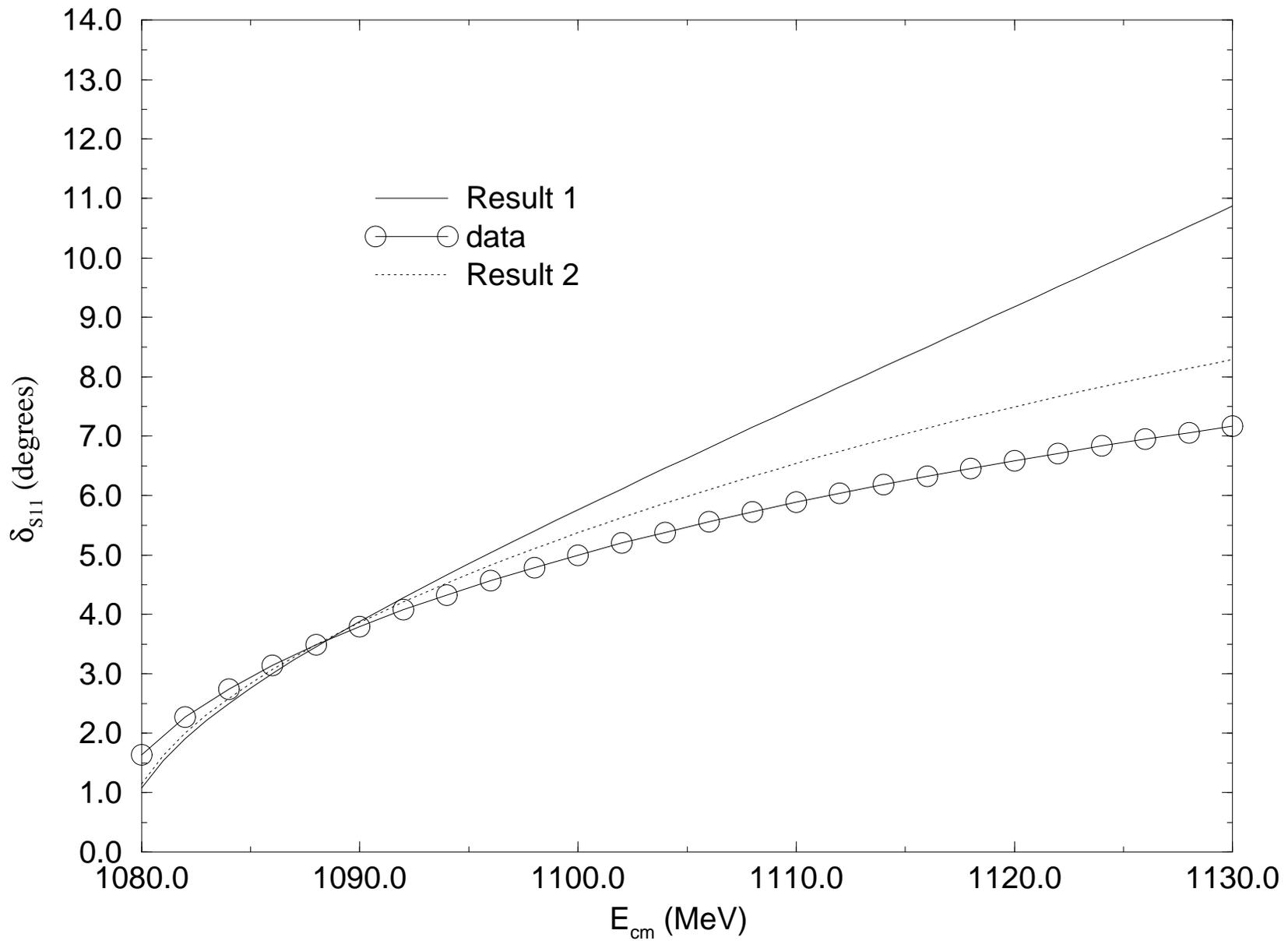
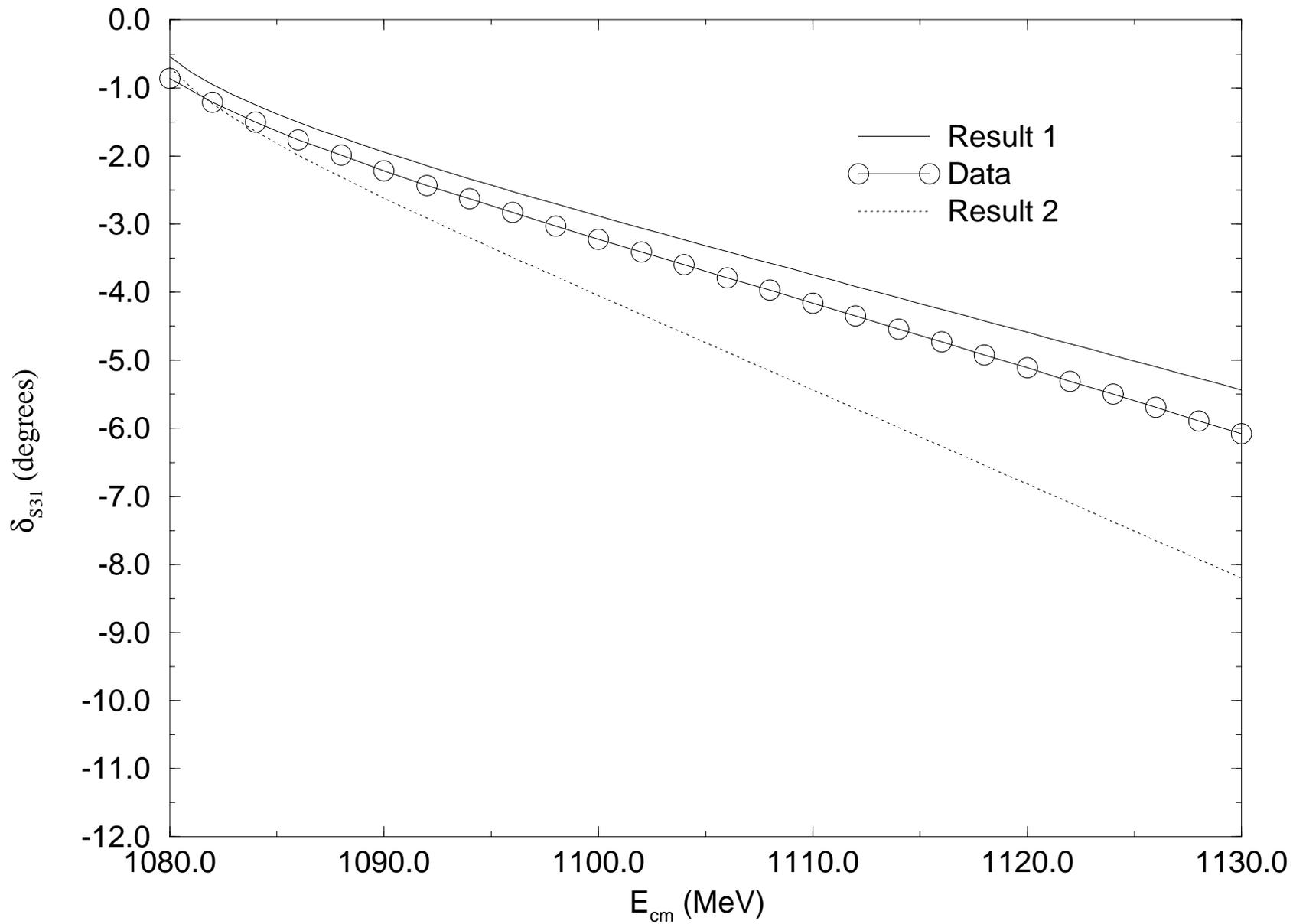
\end{document}